\documentclass[twocolumn,prc,aps,amsmath,amssymb,10pt]{revtex4}
\date{\today}

\usepackage{graphicx}
\usepackage{dcolumn}
\usepackage{float}
\usepackage{threeparttablex}
\usepackage{threeparttable}
\usepackage{longtable,supertabular}
\usepackage{subfigure}
\usepackage[colorlinks=true,allcolors=blue]{hyperref}

\begin{document}

\title{The union 
of rotational and vibrational modes in generator-coordinate-type calculations, with application to neutrinoless double-beta decay}
\author{Changfeng Jiao}
\author{Calvin W. Johnson}
\affiliation{Department of Physics, San Diego State University, San Diego, California 92182-1233, USA}

\begin{abstract}
Good many-body methods for medium and heavy nuclei are important. 
 Here we combine  rotational motion with vibrational modes
in a novel generator-coordinate method (GCM): starting from 
a mean-field solution (Hartree-Fock-Bogoliubov), we create non-orthogonal reference states 
from low-lying quasiparticle Tamm-Dancoff modes, and then project onto states of good angular momentum and particle number. 
The results we benchmark against full shell model calculations.
 Even with just a few such modes we find 
 improvement over standard GCM calculations
in excitation spectra. We also  find significant improvement in $0\nu\beta\beta$ nuclear matrix elements.
\end{abstract}


\maketitle

For  tests of fundamental symmetries such as neutrinoless double-beta ($0\nu\beta\beta$) decay \cite{Avignone08,menendez2017neutrinoless,engel2017status}, and investigations into astroparticle physics
such as direct detection of dark matter \cite{engel1992nuclear,gaitskell2004direct,PhysRevC.89.065501}, one needs accurate interaction matrix elements for medium- to heavy-mass nuclides. 
Despite a  portfolio of available methods~\cite{BG77,Ring04} including  recent developments 
\cite{pieper2005quantum,hagen2010ab,hergert2016medium}, the nuclear many-body problem remains numerically challenging.

Among available  methods, the configuration-interacting shell model (SM) is appealing for its intuitive interpretation, 
straightforward implementation,
and wide usage both phenomenologically \cite{br88,ca05} and \textit{ab initio}  \cite{navratil2000large,barrett2013ab}.  SM calculations boil down to 
diagonalizing in an orthonormal basis a second-quantized (occupation space) Hamiltonian \cite{BG77},
\begin{equation}
\hat{H} = \sum_{ij} \epsilon_{ij} \hat{a}^\dagger_i \hat{a}_j + \frac{1}{4} 
\sum_{ijkl} V_{ijkl}  \hat{a}^\dagger_i  \hat{a}^\dagger_j \hat{a}_l \hat{a}_k ,
\label{ham}
\end{equation}
where $\hat{a}^\dagger_i, \hat{a}_i$ are   fermion  creation and annihilation operators for single-particle states labeled by $i,j,k,l$.
The downside of the SM method, however, 
is that in order to include relevant correlations, one needs a large number of basis states, and the dimension of the many-body 
basis increases exponentially with the number of valence nucleons and/or the number of single-particle states. This limits the 
numerically exact diagonalization of SM Hamiltonian to light nuclei or nuclear models with relatively few valence particles.


 Alternatively, one can build  correlations by diagonalizing the Hamiltonian 
 in a small basis of nonorthogonal states. Here there are 
 two  branches. The first  fall under the rubric of generator coordinate methods (GCM) \cite{Ring04,BenderRMP03}, while the second is known as the Monte Carlo shell model (MCSM)\cite{PhysRevLett.77.3315,otsuka2001monte}.  
  In this paper we  propose a novel approach to the GCM: 
    we combine  using angular momentum projection to extract rotational motion, 
with using low-lying vibrational modes to generate a set of reference states.  Our initial results, applied to $0\nu\beta\beta$ 
transition matrix elements, show promising improvements over common GCM approaches.
 
 The GCM is itself an application of the Hill-Wheeler equation \cite{Ring04},
 $ \int \langle \Psi(\lambda^\prime) | \hat{H} | \Psi(\lambda) \rangle \, d\lambda = E 
 \int \langle \Psi(\lambda^\prime) |  \Psi(\lambda) \rangle \, d\lambda,$
 where $\lambda$ is some parameter which in practice is  
 discretized.  
In GCM 
 the (discretized)  set of $\{ | \Psi(\lambda) \rangle \}$, or reference states,  are typically either particle-number conserving Slater determinants, or quasiparticle vacua, and are typically found 
 by minimizing
\begin{equation}
\langle \Psi(\lambda) | \hat{H} - \lambda \hat{Q} | \Psi (\lambda) \rangle, \label{variational}
\end{equation}
where $\lambda$ 
is now a Lagrange multiplier, and $\hat{Q}$ is some external field, the exact choice of which is part of the art of GCM;
often there are several external fields and corresponding  Lagrange multipliers. 
When using quasiparticles there are  Lagrange multipliers, here suppressed, to constrain the number of protons and neutrons.
Thus one solves constrained Hartree-Fock (HF)  or constrained Hartree-Fock-Bogoliubov (HFB) equations for several different values of 
$\lambda$ (and possibly different $\hat{Q}$) and traditionally one interprets either $\lambda$ or, more commonly, the 
expectation value $\langle \Psi(\lambda) | \hat{Q} | \Psi(\lambda)  \rangle$, as a generalized ``coordinate'' 
that ``generates'' the basis, hence the name. 
Of course, for $\lambda = 0$ one has the original HF or HFB equations. 
It is useful to imagine an energy landscape in the space of Slater determinants or quasiparticle   vacua: the HF or HFB state is at the global minimum, and GCM explores the 
energy landscape around that minimum.   Because these simple states almost always break rotational symmetry, and in the case 
of HFB, particle-number conservation, one needs to restore good angular momentum and particle number through projection.  It is through
the process of projection and then the mixing of multiple projected reference states, that one builds up important correlations. 
Angular momentum projection on deformed states in particular picks out rotational bands, and very often the excitation spectrum 
from a single angular-momentum projected state looks like a rotational band, even if the low-lying spectrum in the full space is not dominated by such bands.


An important question to answer is how to choose the external field(s) $\hat{Q}$ and thus the non-orthogonal reference states. 
 GCM calculation typically focus on  collective correlations known to be important in nuclei.  Most prominent are 
 quadrupole deformations,  originally with axial  symmetry~\cite{RodriguezPRC02,NiksicPRC06_2} but    later extended to   triaxially-deformed configurations~\cite{BenderPRC08,Yao10,RodriguezPRC10,Jiao17}. There are also studies in consideration of octupole modes~\cite{BonchePRL91,YaoPRC15}, as well as investigations of fluctuations in like-particle pairing~\cite{HeenenEPJA01,VaqueroPRL13}. Recent work has indicated the transition operators of double-beta ($\beta\beta$) decay are sensitive to proton-neutron ($pn$) pairing correlations~\cite{VogelPRL86, Engel88,Simkovic08,Mustonen13}.
  The inclusion of $pn$ pairing correlations in  GCM calculations ~\cite{Hinohara14,Javier16,Jiao17,Jiao18} significantly reduces the large difference in the $0\nu\beta\beta$ nuclear matrix elements between previous GCM and SM predictions~\cite{Javier16, Jiao17,Jiao18}. 
 Despite this success, even if one treats both quadrupole deformation and $pn$ pairing as coordinates, and uses the same SM Hamiltonian, the $0\nu\beta\beta$ matrix elements of $^{124}$Sn, $^{130}$Te, and $^{136}$Xe given by GCM are still overestimated by about 30$\%$  when compared with SM results ~\cite{Jiao18}. This implies that some other correlations should not be neglected for an accurate description of $0\nu\beta\beta$ decay, and perhaps other nuclear properties. In addition, previous GCM work neglects, for the most part, non-collective correlations. How to pin down all the collective and non-collective correlations that may play relevant roles is still an open question.   

Instead of choosing the constraining fields by hand, we let the system itself dictate the reference states, through the 
use of a theorem of Thouless ~\cite{Ring04}: the exponential of any one-body operator  $\hat{Z}$
acting on a Slater determinant $| \Psi \rangle$  is another Slater determinant $| \Psi^\prime \rangle$, 
\begin{equation}
\exp( \hat{Z} ) | \Psi \rangle = | \Psi^\prime \rangle \equiv | \Psi(Z) \rangle \label{thouless}.
\end{equation}
Here we use Thouless' theorem as applied to quasiparticle vacua to define an energy landscape  
 $E(Z) = \langle \Psi(Z) | \hat{H} | \Psi(Z) \rangle$,  which can be expanded in $Z$.  
Setting the first derivatives with respect to $Z$ to be zero  yields the Hartree-Fock/HFB equations. 
The second derivatives, which approximate the landscape as a quadratic in $Z$ and thus 
 a multi-dimensional harmonic oscillator, lead to the Tamm-Dancoff and random-phase approximations~\cite{Ring04}, and their quasiparticle 
extensions, for excited states. Closely related to this is the derivation of  the random phase approximation from
a generator coordinate picture \cite{jancovici1964collective}. 

(We note  that the  Monte Carlo shell model\cite{PhysRevLett.77.3315,otsuka2001monte}, itself inspired  by  the imaginary-time-evolution auxiliary-field path-integral approach to the SM\cite{PhysRevC.48.1518,koonin1997shell}, uses 
Thouless' theorem to stochastically explore the energy landscape.  One can relate the path integral, and thus the MCSM, 
to the random phase approximation \cite{PhysRevC.42.R1830}.  While we do not sample the energy surface stochastically, 
the relationship between Thouless' theorem, the energy landscape, the auxiliary-field path integral method, and the MCSM, inspired our development.)

Although one can work in a particle-conserving formalism (indeed, we are exploring this approach separately) for  this 
paper we focus on quasiparticles. In this formalism, detailed below,  excited states are modeled as collective linear
  combinations of two-quasiparticle  excitations, with low-lying collective modes found  by diagonalizing the Hamiltonian 
in this space, also known as  the quasiparticle Tamm-Damcoff approximation (QTDA). 
 Because they arise from a harmonic approximation to the landscape, 
one can conceive of them as \textit{vibrational} modes.

Here is where we branch from the default GCM. Rather than exploring the energy landscape by guessing
(albeit with good reason) the important external fields, we solve the QTDA equations, identify the lowest  
QTDA (vibrational) modes, and  use these modes to generate reference states via Thouless' theorem. We then project out 
states of good angular momentum and particle number and diagonalize the Hamiltonian in this basis. 

This is related to previous work which builds reference states using two-quasiparticle excitations 
\cite{PhysRevC.79.014311,PhysRevC.95.024307}.
However those calculations simply used the lowest-lying two-quasiparticle excitations, whereas we use collective superpositions
of two-quasiparticle excitations as given by QTDA. It is important to note that use of Thouless' theorem is key to introducing 
general linear combinations of two-quasiparticle states. 

In what follows  we first briefly sketch the framework of QTDA-driven GCM, including the QTDA operators, Thouless evolution, symmetry restoration, and configuration mixing. We then use it to compute the low-lying level spectra, reduced $E2$ transition probabilities, and $0\nu\beta\beta$ decay matrix elements, compared with the results given by the SM and the GCM using quadrupole deformations and $pn$ pairing amplitudes as coordinates (denoted as ``CHFB-GCM''). Finally, we summarize our conclusions and suggest future work.

The founding HFB state $| \Phi_0 \rangle$, found by minimizing (\ref{variational})  albeit only with constraints on proton and neutron numbers,
 is a vacuum to quasiparticles $\hat{c}_\alpha(0)$:
  \begin{equation}
 \hat{c}_\alpha(0) = \sum_{\beta}\hat{a}_\beta U^*_{\beta \alpha}(0) + \hat{a}^\dagger_\beta V^*_{\beta \alpha}(0),
 \label{qptransform}
 \end{equation}
 where the argument `0' indicates that $| \Phi_0 \rangle$ is the vacuum relative to these quasiparticles, so that 
 $ \hat{c}_\alpha(0) | \Phi_0 \rangle = 0$. This is important as our reference states will be a set of non-orthogonal vacua.

 Low-lying excited states are approximated as linear combinations of two-quasiparticle excitations
 (the equivalent of 
 particle-hole states in particle-conserving schemes with a Slater determinant): $  \hat{Z}_r 
 | \Phi_0 \rangle$ where
 \begin{equation}{\label{eqn:QTDAoperator}}
\hat{Z}_{r} = \frac{1}{2}\sum_{\alpha \alpha^{\prime}}Z^{r}_{\alpha \alpha^{\prime}}
\hat{c}^\dagger_\alpha(0) \hat{c}^\dagger_{\alpha^\prime} (0).
\end{equation}
 To get the(antisymmetric)  coefficients $Z^{r}_{\alpha \alpha^{\prime}}$ of the QTDA operator, one computes the matrix elements of the Hamiltonian in a basis of two-quasiparticle excited states ,
 \begin{equation}
 {\label{eqn:Amatrix}}
  \begin{aligned}
 A_{\alpha \alpha^\prime, \beta \beta^\prime} 
 &= \langle \Phi_0\vert \lbrack \hat{c}_{\alpha^{\prime}} (0)\hat{c}_{\alpha} (0), \lbrack \hat{H},\hat{c}^\dagger_{\beta} (0)  \hat{c}^\dagger_{\beta^{\prime}} (0)\rbrack\rbrack\vert \Phi_0\rangle 
\end{aligned}
\end{equation}
 The matrix $\mathbf{A}$ is the quasiparticle Tamm-Dancoff (QTDA) matrix, and its expression in terms of the Hamiltonian 
matrix elements   (\ref{ham})  and the quasiparticle transformation (\ref{qptransform}) is  found in the literature \cite{Ring04,suhonen2007nucleons}.
 We  solve 
  \begin{equation}{\label{eqn:QTDAeqn}}
\sum_{\beta \beta^\prime} A_{\alpha \alpha^\prime, \beta \beta^\prime} Z^{r}_{\beta \beta^\prime}= E_{r}^{\text{QTDA}}Z^{r}_{\alpha \alpha^\prime}.
\end{equation}
finding
excitation energies  $E_{r}^{\text{QTDA}}$. 
We approximate the QTDA states by 
applying Thouless' theorem:
\begin{eqnarray}
\vert \Phi_r \rangle & = & 
\exp \big( \lambda \hat{Z}_r  \big)  | \Phi_0\rangle 
\approx  | \Phi_0\rangle  + \lambda\hat{Z}_r  | \Phi_0 \rangle
\notag \\
& = &
\text{exp}\Bigg \{\lambda \frac{1}{2}\sum_{\alpha \alpha^\prime}Z^r_{\alpha \alpha^\prime}\hat{c}^{\dagger}_{\alpha}(0)\hat{c}^{\dagger}_{\alpha^{\prime}}(0) \Bigg\} \vert \Phi_0\rangle.
\label{applyqpthouless}
\end{eqnarray}
While in principle we
could sample different values of $\lambda$ similar to the integration over auxiliary fields in the path integral 
method, we found in 
our current work that a single value of $\lambda = 1$ worked sufficiently well. 
 
 With application of Thouless' theorem we have a new quasiparticle vacuum $| \Phi_r \rangle$,  and we need to file 
  new quasiparticle states $\hat{c}_{\alpha}(r)$ that
annihilate it.
Application of Thouless' theorem with particle-hole operators on a Slater determinant is straightforward\cite{PhysRevC.48.1518}, 
but much less 
so for quasiparticle vacua. We follow the development of \cite{egido1995solution}, defining
 the intermediate operators 
\begin{equation}
\tilde{\tt c}_{\alpha}(r) = \hat{c}_{\alpha}(0)+ \lambda \sum_{\mu}\hat{c}^\dagger_\beta(0) Z^r_{\beta \alpha}
\end{equation}
which satisfy $\tilde{\tt c}_{\alpha}(r)\vert \Phi_r\rangle = 0$. However, they are not proper quasiparticle operators, as they do not satisfy  fermion anti-commutation relations:
\begin{equation}
 \Big \{ \tilde{\tt c}_{\alpha}(r), \tilde{\tt c}^{\dagger}_{\alpha^\prime}(r) \Big \} = \delta_{\alpha \alpha^{\prime}} +  \lambda^2 \sum_{\mu}Z^r_{\mu \alpha}Z^{r\, \ast}_{\mu \alpha^{\prime}},
\end{equation}  
We  regain proper anti-commutation relations by defining
\begin{equation}{\label{eqn:Choleski}}
\mathbf{LL}^{\dagger} =\mathbf{ I} +  \lambda^2 \mathbf{Z}^T\mathbf{Z}^{\ast}. 
\end{equation}
where $\mathbf{I}$ is the unit matrix,  and $\mathbf{L}$ is a lower triangular matrix found easily by a Cholesky decomposition, which is  inverted, and then defining 
\begin{equation}
\hat{c}_{\alpha}(r) = \sum_{\alpha^{\prime}}L^{-1}_{\alpha \alpha^{\prime}}\tilde{\tt c}_{\alpha}(r), 
\end{equation}
which have the desired anti-commutation relations,
\begin{eqnarray}
\Big \{\hat{c}_{\alpha}(r), \hat{c}^{\dagger}_{\alpha^{\prime}}(r)\Big\}   =  \sum_{\mu \mu^{\prime}}L^{-1}_{\alpha \mu}\big [L^{-1}_{\alpha^{\prime}\mu^{\prime}}\big]^{\ast}  \big\{\tilde{\tt c}_{\mu}(r), \tilde{\tt c}^{\dagger}_{\mu^{\prime}}(r)\big\} \notag\\ 
 =  \Big ( \mathbf{L}^{-1} \big (\mathbf{I} +  \lambda ^2\mathbf{Z}^T\mathbf{Z}^{\ast}\big) \big [ \mathbf{L}^{-1} \big]^{\dagger}\Big )_{\alpha \alpha^{\prime}} 
= \delta_{\alpha \alpha^\prime}.
\end{eqnarray}

The resultant Bogoliubov transformation matrix of the transformed state $\vert \Phi_r\rangle$ is given by
${\left( \begin{array}{r}
\hat{c}^{\dagger}(r)\\
\hat{c}(r)
\end{array} 
\right )} = $
\begin{eqnarray}
& 
{\left( \begin{array}{cc}
\big [\mathbf{L}^{-1}\big ]^{\ast} & \lambda \big [\mathbf{L}^{-1} \mathbf{Z}^T\big ]^{\ast}\\
 \lambda \mathbf{L}^{-1}\mathbf{Z}^{T} & \mathbf{L}^{-1}
\end{array} 
\right )}
{\left( \begin{array}{ll}
\mathbf{U}(0)^T & \mathbf{V}(0)^T\\
\mathbf{V}(0)^{\dagger} & \mathbf{U}(0)^{\dagger}
\end{array} 
\right )} {
\left( \begin{array}{l}
\hat{a}^{\dagger}\\
\hat{a}
\end{array} 
\right )} \notag \\
 = & 
{\left( \begin{array}{ll}
\mathbf{U}(r)^T & \mathbf{V}(r)^T\\
\mathbf{V}(r)^{\dagger} & \mathbf{U}(r)^{\dagger}
\end{array} 
\right )} {
\left( \begin{array}{l}
\hat{a}^{\dagger}\\
\hat{a}
\end{array} 
\right )}
\end{eqnarray}
where 
\begin{eqnarray}\label{eqn:newUV}
\mathbf{U}(r) &=& (\mathbf{U}(0)+ \lambda \mathbf{V}(0)^{\ast}\mathbf{Z}^{\ast})\big [ \mathbf{L}^{-1}  \big ]^{\dagger} \\
\nonumber 
\mathbf{V}(r)&=& (\mathbf{V}(0)+ \lambda \mathbf{U}(0)^{\ast}\mathbf{Z}^{\ast}) \ \big [ \mathbf{ L}^{-1} \big ]^{\dagger}
\end{eqnarray}
express the change in the quasiparticle amplitudes.
By choosing various QTDA operators, we can obtain a set of basis states $\vert \Phi_r \rangle$ which are transformed using QTDA operators $\hat{Z}_r$. In the current work, we choose the 15 lowest-energy QTDA phonons, which is a 
computationally tractable choice.
 Since the QTDA evolution further breaks the conservation of particle numbers, we also compute the average proton number $\langle N_Z\rangle$ and neutron number $\langle N_N\rangle$ with the QTDA-transformed wavefunctions, to ensure the chosen QTDA-generated basis states gives the correct proton and neutron numbers approximately. Nonetheless, particle-number projection is required.


Once we have obtained a set of reference states, we project out states of good angular momenta as well as good proton and neutron 
numbers,  $|JMK;NZ;r\rangle\equiv \hat{P}^J_{MK}\hat{P}^N\hat{P}^Z|\Phi_{r}\rangle$. Here $\hat{P}^{\prime}$s projects onto well-defined angular momentum $J$ and its $z$-component $M$, neutron number $N$, and proton number $Z$~\cite{Ring04}. 
Using these states we construct  the Hamiltonian kernel $\mathcal{H}_{MK}^J(r; s)$ and the norm kernel $\mathcal{N}_{MK}^J(r; s)$ are given by:
\begin{equation}
\left.\begin{aligned}
&\mathcal{H}_{MK}^J(r;s) =  \langle\Phi_{r}|  \hat{H} \hat{P}^{J}_{MK} \hat{P}^{N}\hat{P}^{Z}|\Phi_{s} \rangle,\\
&\mathcal{N}_{MK}^J(r;s) = \langle\Phi_{r}|\hat{P}^{J}_{MK}\hat{P}^{N}\hat{P}^{Z}|\Phi_{s}\rangle.
\end{aligned}
\right.
\end{equation}

We solve the generalized eigenvalue problem, which is really solving  the discretized Hill-Wheeler equations~\cite{Ring04}:
\begin{equation}{\label{eqn:HWeq}}
    \sum_{K,s}\Bigl\{\mathcal{H}_{MK}^J(r;s)-E^J_{\sigma}\mathcal{N}^J_{MK}(r;s)\Bigr\}f^{JK}_{s \sigma}=0,
\end{equation}
where $\sigma$ labels the eigenstates.
To solve Eq.(\ref{eqn:HWeq}), we diagonalize the norm kernel $\mathcal{N}$ and use the nonzero eigenvalues and corresponding eigenvectors to construct a set of ``natural states''. Then, the Hamiltonian is diagonalized in the space of these natural states to obtain the GCM states $|\Psi^{J}_{NZ\sigma}\rangle$ (see details in Refs.~\cite{RodriguezPRC10, Yao10}). 
The GCM state  is a superposition of  projected states, 
\begin{equation}{\label{eqn:GCMwf}}
|\Psi^{J,M}_{NZ\sigma}\rangle=\sum_{K,s}f^{JK}_{s\sigma}|JMK;NZ;s\rangle,
\end{equation}
using the weight functions $f^{JK}_{s\sigma}$,  from Eq.~(\ref{eqn:HWeq}).

\begin{figure*}\label{fig1} 
\centering
\includegraphics[width=0.9\textwidth]{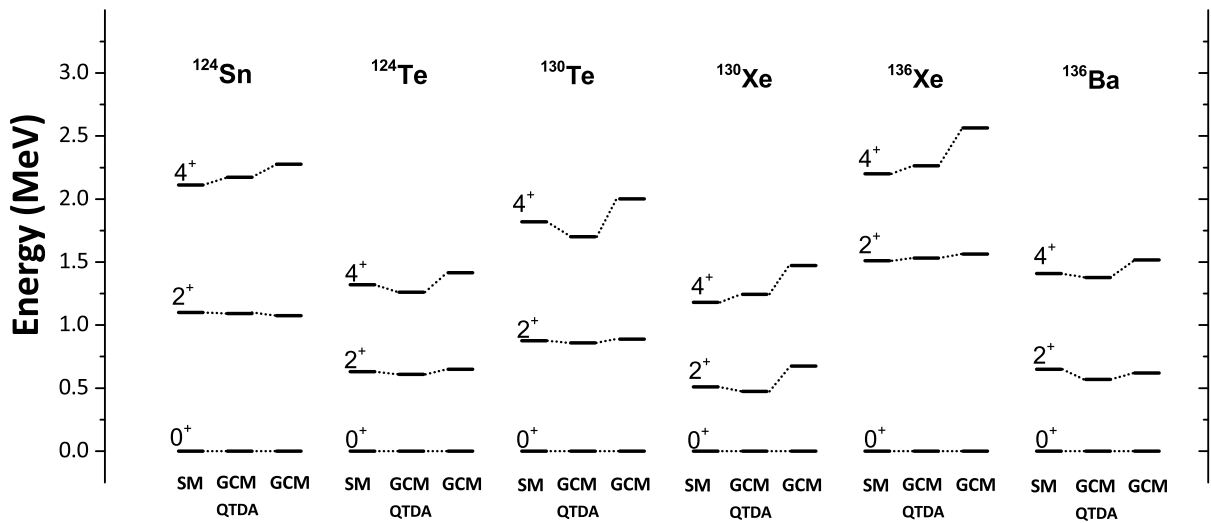}
\caption{The calculated low-lying energy levels for $^{124}$Sn, $^{124}$Te, $^{130}$Te, $^{130}$Xe, $^{136}$Xe, and $^{136}$Ba by using QTDA-driven GCM, compared to the constrained Hartree-Fock-Bogoliubov (CHFB) GCM~\cite{Jiao18} and the exact solutions of SM~\cite{Neacsu15,Horoi16}.}
\end{figure*}

We benchmark our QTDA-driven GCM by comparing results against those from full 
SM calculations, which can be considered numerically ``exact'' in a given model space, as well as a CHFB-GCM calculation, all  
using the same interaction Hamiltonian. 

For this initial work we focus on  the $0\nu\beta\beta$ decay candidate nuclei $^{124}$Sn/Te, $^{130}$Te/Xe, and $^{136}$Xe/Ba. These nuclei are expected to have near spherical or weak deformation. Unlike the well-deformed nuclei which are dominated by rotational behavior, one can expect that the vibrational motion and quasiparticle excitation, which can be described by QTDA, may compete with the rotational motion in the low-lying states of these nuclei. In addition, previous GCM calculations using axial deformation and $pn$ pairing amplitudes as coordinates show about 30$\%$ overestimation of $0\nu\beta\beta$ matrix elements for these nuclei when compared with the SM predictions, implying the lack of important correlations. 


The shell-model effective Hamiltonian we use is called the singular value decomposition (SVD) Hamiltonian~\cite{Qi12}, which is fine tuned for the $jj$55-shell configuration space that compromises the $0g_{7/2}$, $1d_{5/2}$, $1d_{3/2}$, $2s_{1/2}$, and $0h_{11/2}$ orbitals. This effective interaction has accounted successfully for the spectroscopy, electromagnetic transitions, and deformation of the initial and final nuclei that our calculations involve~\cite{Neacsu15,Horoi16}. 

Our CHFB-GCM calculation include axially-symmetric quadrupole deformation and $pn$ pairing amplitude as the coordinates~\cite{Jiao18}. For $^{124}$Te, $^{130}$Te, $^{130}$Xe, and $^{136}$Ba, we generated 50 to 90 basis states, by constraining 9 to 12 different quadrupole moments and 6 to 8 different $pn$ pairing amplitudes, respectively.
The cases  $^{124}$Sn and $^{136}$Xe lack valence protons and valence neutron holes, respectively, so one cannot change $pn$ 
pairing; for those nuclides 
 we use 13 basis states constrained to  quadrupole moments only. It is important to note that as these nuclides have 
 axially symmetric HFB minimua, the standard calculations omit triaxially deformed configurations, so that angular momentum projection only evaluates integrals over the Euler angle $\beta$.  (For other candidate $0\nu\beta\beta$ targets with naturally triaxial minima, e.g. $^{76}$Ge, full  three-dimensional projection unsurprisingly improves results \cite{Jiao17}.)

\begin{table}
\caption{The g.s. energies (in MeV) obtained with SVD Hamiltonian by using CHFB-GCM, QTDA-driven GCM, and SM for $^{124}$Sn, $^{124}$Te, $^{130}$Te, $^{130}$Xe, $^{136}$Xe, and $^{136}$Ba. \label{table_gs_energy}
}
\begin{ruledtabular}
\begin{tabular}{lccc}
 Nuclei & CHFB-GCM & QTDA-GCM& SM \\
\hline
$^{124}$Sn    & $-15.733$  &$-15.684$ &$-16.052$ \\
$^{124}$Te    & $-23.082$ & $-22.641$&$-24.446$  \\
$^{130}$Te    & $-25.705$ & $-25.563$&$-26.039$  \\
$^{130}$Xe    & $-32.583$ & $-32.220$&$-33.313$  \\
$^{136}$Xe    & $-34.931$ & $-34.912$&$-34.971$  \\
$^{136}$Ba    & $-40.341$ & $-40.176$&$-40.745$  
\end{tabular}
\end{ruledtabular}
\end{table}

In our QTDA-driven GCM calculations  we employ about 15 QTDA operators $Z_{r}$  built on the HFB states. 
Aside from the base HFB minimum, we do not use any of the constrained reference states from the CHFB-GCM calculation.
The nuclides $^{124}$Te, $^{130}$Te, $^{130}$Xe, and $^{136}$Ba have axially deformed HFB solutions, but since  $^{124}$Sn and $^{136}$Xe have spherical minima,
we constrain the base HFB states  $| \Phi_0 \rangle$ to axial quadrupole deformation $\beta_2=0$ and $\beta_2=\pm0.1$. 
While both the HFB state and the TDA modes have good axial symmetry, that is, good $J_z$ projection $K$, 
if a QTDA mode has $K \neq 0 $, then exponentiated it mixes $K$. Thus  we must project by quadrature over all three Euler angles.
We also cannot use time-reversal and simplex symmetries which reduce by a factor of 16 the computational requirements of the quadrature 
~\cite{RodriguezPRC10}. Thus, although our current calculations have fewer reference states, the overall computational 
burden is significantly greater, requiring $\sim 2000 \times$ more evaluations for a single matrix element.  We emphasize this difference is largely because our
 CHFB-GCM  code has been highly optimized for nuclides with axially symmetric minima. 
 We have investigated  efficient projection methods \cite{PhysRevC.96.064304,johnson2018convergence} but  have  yet to  implement them in  projected HFB.

 

Figure 1 shows the low-lying level spectra of  $^{124}$Sn, $^{124}$Te, $^{130}$Te, $^{130}$Xe, $^{136}$Xe, and $^{136}$Ba, compared to the CHFB-GCM calculation~\cite{Jiao18}, as well as the SM calculations~\cite{Neacsu15,Horoi16}. Generally, the $2^+$ states obtained from both  CHFB-GCM and QTDA-driven GCM calculations are in reasonable agreement with the SM results and experimental spectra. CHFB-GCM calculations, however, tend to overestimate the excitation energies of $4^+$ states. The overestimation could be due to the fact that the previous GCM calculations exclude the vibrational motion and broken-pair excitation, while these two excitation modes may lower the excited states significantly, especially in the nearly spherical and weakly deformed nuclei. Since the reference states generated by QTDA modes incorporate (at least partly) vibrational motion and two-quasiparticle excitations, our current results  reduce the overestimation of $4^+$ states. Inclusion of the vibrational motion and two-quasiparticle configurations is, indeed, important for a better description of the low-lying spectra of spherical and weakly deformed nuclei. 

Conversely, the ground state energies, shown in Table \ref{table_gs_energy}, are about 50 to 350 keV higher for our current QTDA-driven GCM than for 
the CHFB-GCM, which in turn is about 40 keV to 1.4 MeV above the numerically exact SM energies, a failing for GCM 
calculations previously known \cite{PhysRevC.98.054311}. Keep in mind that our 
QTDA-driven GCM has fewer reference states than CHFB-GCM. 

\begin{table}
\caption{The $B(E2:0_1^+ \rightarrow 2_1^+)$ (in $e^2\text{b}^2$) obtained with SVD Hamiltonian by using CHFB-GCM~\cite{Jiao18}, QTDA-driven GCM (this work), and SM~\cite{Neacsu15, Horoi16} for $^{124}$Sn, $^{124}$Te, $^{130}$Te, $^{130}$Xe, $^{136}$Xe, and $^{136}$Ba, compared to the adopted values~\cite{Pritychenko16}.
\label{tableE2}}
\begin{ruledtabular}
\begin{tabular}{lcccccc}
  & $^{124}$Sn & $^{124}$Te & $^{130}$Te & $^{130}$Xe & $^{136}$Xe & $^{136}$Ba\\
\hline
CHFB-GCM   & 0.168 & 0.648 & 0.165 & 0.492 & 0.220 & 0.475 \\ 
QTDA-GCM & 0.137 & 0.547 & 0.145 & 0.415 & 0.180 & 0.418 \\
SM      &0.146 & 0.579 & 0.153 & 0.502 & 0.215 & 0.479  \\
Adopted & 0.162 & 0.560 & 0.297 & 0.634 & 0.217 & 0.413
\end{tabular}
\end{ruledtabular}
\end{table}

For the calculation of the reduced E2 transition probability $B(E2:0_1^+ \rightarrow 2_1^+)$, we use the canonical effective charges $e^{\text{eff}}_n=0.5e$ and $e^{\text{eff}}_p=1.5e$ for $^{130}$Te/Xe and 136Xe/Ba. For 124Sn/Te, we use $e^{\text{eff}}_n=0.88e$ and $e^{\text{eff}}_p=1.88e$, which are suggested for tin isotopes with no protons in the valence space in Refs.~\cite{Qi12, Horoi16}. The results are compared in Table \ref{tableE2} against SM results~\cite{Neacsu15, Horoi16}, and CHFB- GCM~\cite{Jiao18}, as well as the experimentally adopted values~\cite{Pritychenko16}. We  see a good agreement amongst our QTDA-driven GCM, CHFB-GCM, and SM calculations. Both GCM and SM calculations reproduce well the adopted values with only a slight underestimation for $^{130}$Te and $^{130}$Xe. While the underestimation, also mentioned in Ref.~\cite{Neacsu15}, suggests the effective charges should be adjusted, for 
our purposes we are interested in comparison of the calculations.

As an application of our  QTDA-driven GCM, we compute the values of $0\nu\beta\beta$ decay matrix elements of $^{124}$Sn, $^{130}$Te, and $^{136}$Xe. The results are listed in Table III, where we show the Gamow-Teller, the Fermi, and the tensor contributions respectively. The total matrix elements given by QTDA-driven GCM are smaller than those of CHFB-GCM, in closer agreement with  exact SM diagonalization.

In the closure approximation, one  computes the $0\nu\beta\beta$  matrix element of a two-body transition operator between   the initial and final ground states. Assuming an exchange of a light Majorana neutrino with the usual left-handed currents, the  matrix element is ~\cite{Simkovic08}:
\begin{equation}\label{eqn:ME}
M^{0\nu}=M^{0\nu}_{\text{GT}}-\frac{g^2_V}{g^2_A}M^{0\nu}_{\text{F}} + M^{0\nu}_{\text{T}} \\
\end{equation}
where GT, F, and T refer to the Gamow-Teller, Fermi and tensor parts of the matrix elements. The vector and axial coupling constants are taken to be $g_V=1$ and $g_A =1.254$, respectively. The wave functions are modified at short distances using a Jastrow short-range correlation (SRC) function with CD-Bonn parametrization~\cite{Simkovic08,Simkovic09}. 

The CHFB-GCM agrees well with the SM for the Fermi contribution, but overestimates the Gamow-Teller contribution by 
about 30$\%$~\cite{Jiao18}.  
By contrast, QTDA-driven GCM improves considerably the agreement of the Gamow-Teller contribution, although both Gamow-Teller and Fermi are still slightly overestimated.  This suggests correlations 
beyond quadrupole and $pn$-pairing are important. 

\begin{table}
\caption{The nuclear matrix elements obtained with SVD Hamiltonian by using CHFB-GCM\cite{Jiao18}, QTDA-driven GCM (this work), and SM~\cite{Neacsu15, Horoi16} for $^{124}$Sn, $^{130}$Te, and $^{136}$Xe. CD-Bonn SRC parametrization was used.}
\begin{ruledtabular}
\begin{tabular}{llcccc}
 & & $M^{0\nu}_{\text{GT}}$ &$M^{0\nu}_{\text{F}}$ & $M^{0\nu}_{\text{T}}$ &$M^{0\nu}$ \\
\hline
$^{124}$Sn &CHFB-GCM  &2.48&$-0.51 $&$-0.03$& 2.76\\
                    &QTDA-GCM   &2.08&$-0.73 $&$-0.01$& 2.53\\
                    &SM     &1.85&$-0.47 $& $-0.01$& 2.15\\  
$^{130}$Te &CHFB-GCM  &2.25&$-0.47 $&$-0.02$& 2.52\\
                    &QTDA-GCM   &1.97&$-0.69 $&$-0.01$& 2.39\\
                   &SM     &1.66&$-0.44 $& $-0.01$& 1.94\\  
$^{136}$Xe &CHFB-GCM  &2.17&$-0.32 $&$-0.02$& 2.35\\
                    &QTDA-GCM   &1.65 &$-0.50 $&$-0.01$& 1.96\\
                    &SM    &1.50&$-0.40 $& $-0.01$& 1.76 
\end{tabular}   
\end{ruledtabular}
\end{table}


In this paper, we present a new Hamiltonian-based GCM calculation. 
In this approach, we include low-lying vibrational modes, derived from QTDA, to evolve the HFB minimum via Thouless' theorem to generate references states. These references states are then projected to have good particle number and angular momentum. 
To show the reliability of the new QTDA-driven GCM calculation, we apply it to weakly deformed nuclides and $0\nu\beta\beta$ decay candidate nuclei pairs $^{124}$Sn/Te, $^{130}$Te/Xe, and $^{136}$Xe/Ba, and compared 
 the low-lying level spectra, the reduced $E2$ transition probabilities $B(E2:0_1^+ \rightarrow 2_1^+)$, and the $0\nu\beta\beta$ decay matrix elements, against  exact diagonalization SM results, as well as CHFB-GCM calculations based on the HFB states constrained to different amount of collectivity (i.e., deformation, isovector pairing, isoscalar pairing, \textit{etc}.), all using the same interaction Hamiltonian. Our QTDA-driven GCM results are in reasonable agreement with the SM,  competitive with and in some cases outperforming  CHFB-GCM calculations. 

In the near future we will investigate the efficiency of QTDA-driven GCM; for example,  $K$-mixing by TDA modes requires 
more burdensome computation, and it is worth to see  if it is 
practical to constrain it. In addition, we see two avenues for further   improvements. The first is to add more reference states. Furthermore, one could evolve  HFB states constrained to quadrupole deformation, isoscalar pairing correlation, \textit{etc}., 
or simply combine 
the reference states of CHFB- and QTDA-driven GCM; the results of Tables \ref{table_gs_energy} and \ref{tableE2} suggests this may be a useful or even necessary strategy.  
The other option is, instead of QTDA, one could try quasiparticle random phase approximation (QRPA) operators, which would incorporate two-particle two-hole components into the ground state.  While more complicated, this could  improve the description of the ground state, and hence  a better description $0\nu\beta\beta$ decay nuclear matrix elements needed to interpret experiments.

This material is based upon work supported by the U.S. Department of Energy, Office of Science, Office of Nuclear Physics, 
under Award Number  DE-FG02-03ER41272, and by  U.S. Department of Energy Topical Collaboration Grant No. DE-SC0015376,
and of the National Energy Research Scientific Computing Center, a DOE Office of Science User Facility supported by the Office of Science of the U.S. Department of Energy under Contract No. DE-AC02-05CH11231

\bibliographystyle{apsrev4-1}
\bibliography{0vbbref}

\end{document}